\documentclass[english]{article}
\usepackage[T1]{fontenc}
\usepackage[latin9]{inputenc}
\usepackage{geometry}
\usepackage{amssymb}
\usepackage{graphicx}
\usepackage{setspace}
\usepackage{esint}

\makeatletter
\newcommand{\lyxaddress}[1]{
\par {\raggedright #1
\vspace{1.4em}
\noindent\par}
}

\makeatother

\usepackage{babel}
\begin{document}
\begin{doublespace}

\title{Holography, large scale structure, supermassive black holes and minimum
stellar mass}
\end{doublespace}

\begin{doublespace}

\author{T.R. Mongan}
\end{doublespace}

\maketitle
\begin{doublespace}

\lyxaddress{84 Marin Avenue, Sausalito, California 94965 USA; tmongan@gmail.com}
\end{doublespace}
\begin{abstract}
\begin{doublespace}
This analysis considers our universe as a closed Friedmann universe,
dominated by vacuum energy in the form of a cosmological constant,
with cosmological parameters obtained from full mission \emph{Planck}
satellite observations. A few simple assumptions lead to straightforward
calculation of general features of large scale structures in the universe
and minimum stellar mass as a function of redshift. Those assumptions
also generate upper and lower bounds on supermassive black hole mass
in relation to total stellar mass of the host galaxy, consistent with
observations across four orders of magnitude of black hole mass and
five orders of magnitude of galactic stellar mass. The results are
based only on fundamental constants and measured cosmological parameters.
No arbitrary parameters are involved.\end{doublespace}

\end{abstract}
\begin{doublespace}

\section{Holography in the universe}
\end{doublespace}

\begin{doublespace}
Full mission 2015 \emph{Planck} satellite observations \cite{key-1}
indicate our universe is dominated by vacuum energy, spatially flat
to a good approximation, with Hubble constant $H_{0}=67.8$ km sec$^{-1}$Mpc$^{-1}$,
total matter density $\Omega_{m}=0.308,$ and baryonic density $\Omega_{b}=0.048$.
Accordingly, this analysis treats our universe as a closed Friedmann
universe, dominated by vacuum energy in the form of a cosmological
constant and so large that it is approximately flat. In what follows,
$\rho_{r}(z)$ is the cosmic microwave background (CMB) radiation
density at redshift $z$, where $\rho_{r}(z)=(1+z)^{4}\rho_{r}(0)$
and the mass equivalent of today's radiation energy density $\rho_{r}(0)=4.4\times10^{-34}$g/cm$^{3}$
\cite{key-2}. Correspondingly, $\rho_{i}(z)$ is the matter density
within large scale structure level $i$ at redshift $z$ and $\rho_{0}(0)$
is today's matter density in the universe as a whole. With Hubble
constant $H_{0}=67.8$ km sec$^{-1}$Mpc$^{-1}$, the critical density
$\rho_{crit}=\frac{3H_{0}^{2}}{8\pi G}=8.64\times10^{-30}$g/cm$^{3}$,
where $G=6.67\times10^{-8}$ cm$^{3}$g$^{-1}$sec$^{-2}$ and $c=3.00\times10^{10}$cm
sec$^{-1}$. Since matter accounts for 30.8\% of the energy in today's
universe, $\rho_{0}(0)=0.308\rho_{crit}=2.66\times10^{-30}$g/cm$^{3}$
and the vacuum energy density $\rho_{v}=\left(1-0.308\right)\rho_{crit}=5.98\times10^{-30}$g/cm$^{3}$.
The cosmological constant $\Lambda=\frac{8\pi G\rho_{v}}{c^{2}}=1.12\times10^{-56}$cm$^{2}$
and there is an event horizon in the universe at radius $R_{H}=\sqrt{\frac{3}{\Lambda}}=1.64\times10^{28}$cm.
According to the holographic principle \cite{key-3}, the number of
bits of information available on the light sheets of any surface with
area $a$ is $\frac{a}{4\delta^{2}ln\left(2\right)}$, where $\delta=\sqrt{\frac{\hbar G}{c^{3}}}$
is the Planck length and $\hbar=1.05\times10^{-27}$g cm$^{2}$/sec
is Planck's constant. So, only $N=\frac{\pi R_{H}^{2}}{\delta^{2}ln\left(2\right)}=4.69\times10^{122}$
bits of information on the event horizon will ever be available to
describe our universe. 

In a closed universe, there is no source or sink for information outside
the universe, so the total amount of information available to describe
the universe remains constant. Also, after the first few seconds of
the life of the universe, energy exchange between matter and radiation
is negligible compared to the total energy of matter and radiation
separately \cite{key-4}. Therefore, in a closed universe, the total
quantity of matter in the universe is conserved, there is only a fixed
amount of information available, and the average mass per bit of information
is constant. In a closed, isotropic, and homogeneous Friedmann universe,
the constant mass per bit of information (the mass $M_{H}=\frac{4}{3}\pi R_{H}^{3}\rho_{0}(0)=4.92\times10^{55}$g
within the event horizon today divided by the number of bits of information
within the event horizon) is $\left(4.92\times10^{55}g\right)/\left(4.69\times10^{122}\right)=1.050\times10^{-67}$g.
So, the total mass within the event horizon today relates to the square
of the event horizon radius by $M_{u}=fR_{H}^{2}$, where $f=0.183$
g/cm$^{2}$, giving the relation between mass within the event horizon
and radius of a holographic screen just enclosing that mass. 

This analysis addresses equilibrium conditions of large scale structure
at $z\lesssim6$, but does not address the important details of large
scale structure collisions and mergers accompanying development of
large scales structure as time passes.
\end{doublespace}

\begin{doublespace}

\section{Assumptions about large scale structure}
\end{doublespace}

\begin{doublespace}
A hierarchical self-similar description of large scale structure in
the universe results from three assumptions:
\end{doublespace}
\begin{enumerate}
\begin{doublespace}
\item All information about an isolated gravitationally bound astronomical
structure of mass $M$ is on the light sheets of a holographic spherical
screen with radius $R=\sqrt{\frac{M}{0.183}}$ cm around the center
of mass of the structure, and those bits of information (and the matter
within the screen) are in thermal equilibrium with the CMB radiation.
\item Structures at any given self-similar structural level range in mass
from the Jeans' mass at that level down to the Jeans' mass for the
next finer level of structure.
\item The number of structures of mass $m$ within a structural level is
$\frac{K}{m}$, where $K$ is constant, so the amount of information
in any mass bin (proportional to $\frac{K}{m}m$) is the same in all
mass bins. \end{doublespace}

\end{enumerate}
\begin{doublespace}
The relation between supermassive black hole mass and total mass of
the associated large scale structure is estimated based on two assumptions:
\end{doublespace}
\begin{enumerate}
\begin{doublespace}
\item The supermassive black hole inhabits a core volume within the isothermal
halo of dark matter surrounding the large scale structure, and the
core radius is determined by the holographic radius of sub-elements
of the structure that can maintain circular orbits around the black
hole without being disrupted.
\item Almost all matter in the universe is within the holographic screens
surrounding large scale structures, so the baryon fraction of matter
within the holographic screens at various structural levels is the
same as baryon fraction for the universe as a whole.\end{doublespace}

\end{enumerate}
\begin{doublespace}
No further assumptions are required to estimate minimum stellar mass
as a function of redshift, and none of the following calculations
involve any free parameters.
\end{doublespace}

\begin{doublespace}

\section{Large scale structure at z = 0}
\end{doublespace}

\begin{doublespace}
This analysis identifies three levels of self-similar large scale
structure larger than stellar systems (corresponding to bound superclusters,
galaxies, and star clusters) within the event horizon today. Those
self-similar large scale structures are gravitationally-bound systems
of $n$ widely separated units of the next lower structural level
in a sea of cosmic microwave background photons.

In this analysis, today's speed of pressure waves affecting matter
density at structural level $i$ is $c_{si}(0)=\frac{2c}{3}\sqrt{\frac{\rho_{r}(0)}{\rho_{i}(0)}}$
\cite{key-5}, and the corresponding Jeans' length $L_{i+1}(0)=c_{si}(0)\sqrt{\frac{\pi}{G\rho_{i}(0)}}$
\cite{key-5}. In today's universe, $c_{s0}=2.58\times10^{8}$cm/sec,
and the first level (bound supercluster) Jeans' length $L_{1}(0)=1.09\times10^{27}$cm.
The first level Jeans' mass, the mass of matter within a radius one
quarter of the Jeans' wavelength $L_{1}(0)$, is $M_{1}(0)=\rho_{0}(0)\frac{4}{3}\pi\left(\frac{L_{1}(0)}{4}\right)^{3}=2.24\times10^{50}$g.
All scales smaller than the Jeans' wavelength are stable against gravitational
collapse, and the radius of the spherical holographic screen for the
first level Jeans' mass is $R_{1}=3.50\times10^{25}$cm. The matter
density within the spherical holographic screen for the first level
Jeans' mass is $\rho_{1}(0)=\frac{0.183R_{1}^{2}}{\frac{4}{3}\pi R_{1}^{3}}=1.25\times10^{-27}$g/cm$^{3}$.
Then, $c_{s1}=1.19\times10^{7}$cm/sec within the first level Jeans'
mass, the second level (galaxy) Jeans' length is $L_{2}(0)=2.32\times10^{24}$cm,
and the second level Jeans' mass is $M_{2}(0)=\rho_{1}(0)\frac{4}{3}\pi\left(\frac{L_{2}(0)}{4}\right)^{3}=1.02\times10^{45}$g.
Continuing, the third level (star cluster) Jeans' mass $M_{3}(0)=4.64\times10^{39}$g,
the fourth level (stellar system) Jeans' mass $M_{4}(0)=2.11\times10^{34}$g,
and $\frac{M_{1}(0)}{M_{H}}=\frac{M_{2}(0)}{M_{1}(0)}=\frac{M_{3}(0)}{M_{2}(0)}=\frac{M_{4}(0)}{M_{3}(0)}=4.6\times10^{-6}$.
The hierarchy of large scale structure stops with star clusters, because
stellar systems cannot be treated as $n$ widely separated sub-elements
in a sea of cosmic microwave background photons.

The range of large scale structure masses indicated by this analysis
compares to astrophysical data as follows. The mass of bound superclusters
should be below the first level Jeans' mass, $2.24\times10^{50}$g.
This first level Jeans' mass is about midway between the upper bound
$3.4\times10^{50}$g and the lower bound $1.7\times10^{49}$g estimates
\cite{key-6} of the mass of the Corona Borealis bound supercluster,
one of the largest gravitationally bound structures identified to
date. The upper limit on stellar mass is about $300M_{\odot}$\emph{
}\cite{key-7} and the lower limit is $\approx0.08M_{\odot}$\cite{key-8}.
Kroupa \cite{key-9} estimated the number of stars of mass $m$ in
the range $0.08M_{\odot}$ to $0.5M_{\odot}$ as $\sim m^{-1.3}$
and the number for $m>0.5M_{\odot}$ as $\sim m^{-2.3}.$ So, with
a $300M_{\odot}$upper limit on stellar mass, the 4th level Jeans'
mass at $z=0$ is greater than the mass of 99\% of stars and the 4th
level Jean's mass is a reasonable representation of the mass of the
largest stellar systems.

Identifying bound superclusters as structures with masses between
the first and second level Jeans' masses, galaxies as structures with
masses between the second and third level Jeans' masses, and star
clusters as structures with mass between the third and fourth level
Jeans' masses, the universe within the event horizon today can be
considered successively as an aggregate of bound superclusters, an
aggregate of galaxies, an aggregate of star clusters, or an aggregate
of stellar systems. The Jeans' masses identify each structural level,
but a mass distribution is needed to estimate the number of entities
in each structural level and the average mass of structures at that
level. If the number of structures with mass $m$ within a structural
level is $\frac{K}{m}$ , the number of bound superclusters within
the event horizon is $n=\intop_{4.6\times10^{-6}M_{1}}^{M_{1}}\left(\frac{K}{m}\right)dm=12.3K$
and the mass within the event horizon relates to the aggregate of
bound supercluster masses by $M_{H}=\intop_{4.6\times10^{-6}M_{1}}^{M_{1}}m\left(\frac{K}{m}\right)dm\approx KM_{1}$.
So, $K=\frac{M_{H}}{M_{1}}$, the average mass of a bound supercluster
$\overline{M_{1}}=\frac{M_{H}}{n}=\frac{M_{1}}{12.3}=1.8\times10^{49}$g
and the mass within the event horizon is the number of bound superclusters
times the average bound supercluster mass. There are $n=\intop_{4.6\times10^{-6}M_{2}}^{M_{2}}\left(\frac{K}{m}\right)dm=12.3K$
galaxies in a first level Jeans' mass, and the first level Jeans'
mass is the aggregate of the galaxy masses within that Jeans' mass,
so $M_{1}=\intop_{4.6\times10^{-6}M_{2}}^{M_{2}}m\left(\frac{K}{m}\right)dm\approx KM_{2}$.
Then, $K=\frac{M_{1}}{M_{2}}$, and the average galaxy mass $\overline{M_{2}}=\frac{M_{1}}{n}=\frac{M_{2}}{12.3}=8.3\times10^{43}$g.
A similar analysis gives an average star cluster mass of $3.8\times10^{38}$g,
and these results are consistent with observations \cite{key-10,key-11}. 

Down to the third (star cluster) structural level, the total number
$n=12.3K=2.7\times10^{6}$ of next lower level substructures inside
the holographic screens for the Jeans' length at each structural level
is the same as the total number of bound superclusters within the
event horizon. Furthermore, there are $2.2\times10^{5}$ average mass
galaxies in an average mass bound supercluster and $2.2\times10^{5}$
average mass star clusters in an average mass galaxy. To understand
the self-similarity (scale invariance) of large scale structures,
consider gravitationally-bound systems of $n$ entities with mass
$m$ and total mass $M=nm$. For structures with $n\approx10^{5}$,
the substructure mass $m$ is much less than the mass $M$ of the
next highest level of structure. From the virial theorem, the gravitational
potential energy of the systems is $V_{G}=-\frac{GM^{2}}{2R}.$ If
the information describing gravitationally-bound astronomical systems
of total mass $M$ consisting of $n$ smaller entities with mass $m\ll M$
is available on a spherical holographic screen of radius $R=\sqrt{\frac{M}{0.183}}$
surrounding the system, the gravitational potential energy of the
structure of mass $M$ within the holographic screen is $V_{G}=-\frac{GM^{2}}{2R}=-\frac{G(0.183)^{2}R^{3}}{2}$.
So, self-similarity (scale invariance) of large scale structures occurs
because the average gravitational potential energy per unit volume
at each structural level depends only on the gravitational constant
and is identical for all levels of large scale structure.
\end{doublespace}

\begin{doublespace}

\section{Minimum stellar mass as a function of redshift }
\end{doublespace}

\begin{doublespace}
Stellar systems are the basic elements of self-similar large scale
structures (star clusters, galaxies, bound superclusters, and the
universe within the event horizon), and formation of the first stellar
systems depended on thermonuclear reactions between (strongly interacting)
protons in the baryon fraction of the matter density in the universe.
The mass of the smallest gravitationally bound systems that are stellar
systems at redshift $z$ is estimated by setting the escape velocity
of protons on the holographic screen for the minimum mass stellar
system, with radius $R_{min}$, equal to the average velocity of protons
in equilibrium with CMB radiation outside the screen. For $R>R_{min}$,
the escape velocity (escaping proton temperature) on the holographic
screen is such that escaping protons are at higher temperature than
the CMB and can transfer heat (and energy) to the CMB. Correspondingly,
for $R<R_{min},$ the escape velocity (escaping proton temperature)
on the holographic screen is such that escaping protons would be at
lower temperature than the CMB and unable to transfer heat (and energy)
to the CMB. Protons in equilibrium with the CMB that outside the holographic
screen for systems with mass less than the minimum stellar mass can
transfer heat (and energy) to those structures until they reach the
minimum stellar mass.

The escape velocity for a proton of mass $m_{p}$ gravitationally
bound at radius $R$ from the centroid of a structure with mass $M$
is calculated from $\frac{1}{2}m_{p}v^{2}=\frac{GMm_{p}}{R}$. If
the escape velocity of a proton on the holographic screen for the
minimum mass stellar system at redshift $z$ is the velocity of a
proton in thermal equilibrium with the CMB, $\frac{3}{2}kT=\frac{GMm_{p}}{R}$,
where the CMB temperature $T=(1+z)2.725^{o}K$ and the Boltzmann constant
$k=1.38\times10^{-16}$(g cm$^{2}$/sec$^{2}$)/$^{o}K$. Since the
radius $R$ of the holographic screen for a structure of mass $M$
is $R=\sqrt{\frac{M}{0.183}}$, the minimum mass of stellar systems
at redshift $z$ is $M_{stellar}=\frac{1}{0.183}\left(\frac{1.5k(1+z)2.725}{Gm_{p}}\right)^{2}$.
If outgoing protons near the holographic screen are in thermal equilibrium
with outgoing photon flow from the minimum mass star, a star must
have mass at or above the minimum stellar mass for the system to appear
as a star against the CMB background. Note that radii of holographic
screens for stellar systems are considerably larger than radii of
stars themselves. For example, the radius of the holographic screen
for our sun is comparable to the radius of the entire solar system
including the Oort cloud.

The maximum stellar mass of $300M_{\odot}$ \cite{key-7} coincided
with the minimum stellar mass at $z\approx64$, consistent with indications
that the first stars formed at $z\approx65$ \cite{key-12}. Today,
at $z=0$, the analysis indicates the smallest stellar systems have
masses $>0.07M_{\odot}$, consistent with the mass of the smallest
stars \cite{key-8}. That the holographic principle provides a lower
bound on stellar mass using only the Boltzmann constant, CMB temperature,
$G$, and $m_{p}$ suggests a unifying relation between the organization
of information and the four basic forces (gravity, electromagnetism,
strong interactions, and weak interactions) underlying the relations
embodied in specific equations modeling details of thermonuclear reactions
and stellar dynamics. That idea gains further support from the fact
that the 4th level Jeans' mass at $z=0$, estimating the upper bound
on stellar system mass, is greater than the 99th percentile mass of
stars in Kroupa's approximate mass distribution.
\end{doublespace}

\begin{doublespace}

\section{Large scale structure at z > 0}
\end{doublespace}

\begin{doublespace}
At redshift $z>0$, when the matter density $\rho_{0}(z)$ is much
greater than the radiation density $\rho_{r}(z)$, the speed of pressure
waves affecting matter density at redshift $z$ within structural
level $i$ is $c_{si}(z)=c\sqrt{\frac{4(1+z)^{4}\rho_{r}(0)}{9\rho_{i}(z)}}$
\cite{key-7}, and the Jeans' length at that level $L_{i+1}(z)=c_{si}(z)\sqrt{\frac{\pi}{G(1+z)^{3}\rho_{i}(z)}}$
\cite{key-7}. The first level of large scale structure within the
universe is determined by the Jeans' mass $M_{1}(z)=\frac{4\pi}{3}\left(\frac{L_{1}(z)}{4}\right)^{3}\rho_{0}(z)$,
where $L_{1}(z)=\frac{\left(1+z\right)^{2}}{\rho_{0}(z)}\frac{2c}{3}\sqrt{\frac{\pi\rho_{r}(0)}{G}}=\frac{(1+z)^{2}B}{\rho_{0}(z)}$.
Since $B=\frac{2c}{3}\sqrt{\frac{\pi\rho_{r}(0)}{G}}$ is independent
of $z$, the first level Jeans' mass $M_{1}(z)=M_{1}=\frac{\pi B^{3}}{48\rho_{0}^{2}(0)}$
is independent of $z$ \cite{key-7}. Evolution of large scale structure
is characterized by $N(z)$, the number of structural levels between
the Jeans' mass $M_{1}$ and stellar systems, and $n(z)$, the average
number of next lower level structures within a structure at any given
level, as structures in the $N(z)$ levels coalesce into the three
levels present today. The Jeans' mass $M_{i}(z)$ of structures in
level $i$ is determined by the Jean's length $L_{i}(z)$ in that
structural level and the holographic density $\rho_{i-1}(z)$ inside
the holographic screen for the Jeans' mass $M_{i-1}(z)$ of the next
highest structural level. So, the ratio of the Jeans' mass $M_{i}(z)$
to the Jeans' mass $M_{i+1}(z)$ in the next subordinate level is
$\frac{M_{i}(z)}{M_{i+1}(z)}=\frac{L_{i-1}^{3}(z)\rho_{i-1}(z)}{L_{i}^{3}(z)\rho_{i}(z)}=\frac{\rho_{i}^{2}(z)}{\rho_{i-1}^{2}(z)}$.
The holographic density $\rho_{i}(z)=\frac{3A}{4\pi R_{i}(z)},$where
$A=0.189\frac{g}{cm^{2}}$ and the radius of the holographic screen
for the Jeans' mass $M_{i}(z)$ is $R_{i}(z)=\sqrt{\frac{\pi B^{3}(1+z)^{6}}{48A\rho_{i}^{2}(z)}}.$
So, $\frac{M_{i}(z)}{M_{i+1}(z)}=\frac{\rho_{i}^{2}(z)}{\rho_{i-1}^{2}(z)}=\left(\frac{3A}{\pi B}\right)^{3}\frac{1}{(1+z)^{6}}=\frac{2.7\times10^{6}}{(1+z)^{6}}$.
If the number of structures $n\left(m\right)$ in mass bin $m$ is
$n\left(m\right)=\frac{K}{m}$, the average mass $\overline{M_{i}(z)}$
of structures in level $i$ is the total mass of the next lowest level
of structures within level $i$ divided by the total number of next
lowest level of structures within level $i$. So, $\overline{M_{i}(z)}=\left(\int_{M_{i+1}(z)}^{M_{i}(z)}m\frac{K}{m}dm\right)/\left(\int_{M_{i+1}(z)}^{M_{i}(z)}\frac{K}{m}dm\right)=M_{i}(z)\left(1-\frac{M_{i+1}(z)}{M_{i}(z)}\right)/\left(\ln\left(\frac{M_{i}(z)}{M_{i+1}(z)}\right)\right)$.
The number $n(z)$ of average mass structures of next lower level
within the average mass at any structural level is $n(z)=\frac{\overline{M_{i}(z)}}{\overline{M_{i+1}(z)}}=\frac{M_{i}(z)}{M_{i+1}(z)}=\left(\frac{3A}{\pi B}\right)^{3}\frac{1}{\left(1+z\right)^{6}}=\frac{2.7\times10^{6}}{(1+z)^{6}}$
and the number $N(z)$ of self-similar structural levels exceeding
the minimum stellar system mass $M_{min\, stellar}(z)$ is the integer
truncation of $\frac{1}{log(\frac{M_{i}}{M_{i+1}})}log(\frac{M_{1}}{M_{min\, stellar}(z)})$.
Since $n\left(z\right)$ must be greater than 2 in a hierarchical
model of large scale structure, the hierarchical analysis above is
inappropriate at $z>5.92$ and probably not appropriate until $n(z)>10$
at $z<4.29$, when the analysis indicates sixteen self-similar structural
levels.

Three other comparisons related respectively to the average masses
of bound superclusters, galaxies and star clusters are worth considering.
For bound superclusters, combining the virial theorem with the holographic
relation $M=0.183R^{2},$ the average root mean square velocity of
subelements in a self-similar large scale structure of mass $M$ is
$v_{rms}=\sqrt{\frac{G}{2}}\left(0.183M\right)^{\frac{1}{4}}$. The
closing velocity of the colliding ``bullet cluster'' galaxies 1E0657-56
\cite{key-13} at $z=0.3$ is estimated at $4.8\times10^{8}$cm/sec,
roughly twice the r.m.s galaxy velocity of $2.6\times10^{8}$cm/sec
estimated for the average $z=0.3$ bound supercluster mass $2.1\times10^{49}$g. 

Second, the holographic principle relates mass and angular momentum
of large scale structures, as found by Wesson \cite{key-14}. If large
scale structures exist within isothermal spherical halos with $\frac{1}{r^{2}}$
density distributions, the angular momentum of large scale structures
is $J=I\omega$, where the moment of inertia $I$ of an isothermal
spherical system of mass $M$ is $I=\frac{2}{9}MR^{2}$, and $\omega$
is the angular velocity of the system. Using the holographic relation
$M=0.183R^{2}$ yields $J=\left(\frac{2}{9}\right)\left(\frac{M^{2}}{0.183}\right)\omega$.
The angular velocity is estimated by considering a mass $m$ fixed
on the surface of the rotating structure just inside the holographic
screen for the structure, with radius $R_{s}$. The radial acceleration
of that particle $a_{r}=-\omega^{2}R_{s}$ results from the gravitational
force $F_{r}=-\frac{GmM}{R_{s}^{2}}$ attracting the particle to the
centroid of the structure, so $\omega^{2}=\frac{GM}{R_{s}^{3}}=\frac{G}{\sqrt{0.183M}}$.
The result is $J=p(M)M^{2}=\frac{2}{9}\frac{G^{0.5}}{(0.183M)^{0.25}}M^{2}$.
Then, $p(M)=9\times10^{-16}$ for an average galactic mass of $8.3\times10^{43}$g,
15\% higher than Wesson's empirical value $p=8\times10^{-16}$\cite{key-14}.

Third, Forbes and Kroupa \cite{key-15} suggest galaxies and star
clusters have different relaxation times, with galaxy relaxation times
greater than the age of the universe and star cluster relaxation times
similar to the age of the universe. Based on standard texts (Shu \cite{key-16}
and Binney \& Tremaine \cite{key-17}), Bhattacharya \cite{key-18}
considers systems of mass $M$ and radius $R$ composed of $N$ elements
with average mass $m$ and number density $n=\frac{3N}{4\pi R^{3}}$
and approximates the two body relaxation time for those system as
$t_{R}\approx\frac{0.1N}{\ln N\sqrt{Gmn}}$. Using the holographic
relation $R=\sqrt{\frac{M}{0.183}}$ between mass and radius of a
system, its relaxation time is $t_{R}\approx\frac{0.1}{\ln N}\sqrt{\frac{4\pi N}{3Gm}}\left(\frac{M}{0.183}\right)^{\frac{3}{4}}$.
The above analysis indicates today's average masses of bound superclusters,
galaxies and star clusters are, respectively, $2.1\times10^{49}$g,
$8.3\times10^{43}$g, and $3.8\times10^{38}$g. If average stellar
mass is about the solar mass, the relaxation time for an average mass
star cluster is about $6\times10^{17}$sec, comparable to the age
of the universe at $13.6\times10^{9}$yr $=4.29\times10^{17}$sec.
In contrast, consistent with Forbes and Kroupa \cite{key-15}, relaxation
times for average mass galaxies and bound superclusters are $1\times10^{19}$sec
and $3\times10^{20}$sec, considerably longer than the age of the
universe. 
\end{doublespace}

\begin{doublespace}

\section{Supermassive black holes}
\end{doublespace}

\begin{doublespace}
If visible large scale structures develop within isothermal spherical
halos of dark matter, the matter density distribution in large scale
structures is approximated by $\rho(r)=\frac{a}{r^{2}}$, where $r$
is the distance from the center of the structure and $a$ is constant.
In this regard, Pato and Iocco \cite{key-19} did a non-parametric
reconstruction of the dark matter profile of our galaxy directly from
observations. Their results indicate an isothermal profile fits observations
at least as well as other commonly used profiles.

The mass $M_{s}$ within the holographic radius $R_{s}$ in an isothermal
density distribution is $M_{s}=4\pi\int_{0}^{R_{s}}\frac{a}{r^{2}}r^{2}dr=4\pi aR_{s}$,
requiring $a=\frac{M_{s}}{4\pi R_{s}}$. Since the mass within radius
$R$ from the center of a large scale structure is $M_{R}=4\pi\int_{0}^{R}\frac{a}{r^{2}}r^{2}dr=\frac{R}{R_{s}}M_{s}$,
the tangential speed $v_{t}$ of a sub-element of mass $m$ in a circular
orbit of radius $R$ around the center is found from $\frac{GMm}{R^{2}}=\frac{4\pi Gam}{R}=\frac{mv_{t}^{2}}{R}$.
So, the tangential speed of sub-elements in circular orbits around
the center, $v_{t}=\sqrt{G\frac{M_{s}}{R_{s}}}$, does not depend
on distance from the center and sub-elements tend to lie on a flat
tangential speed curve. With an $\frac{a}{r^{2}}$ matter density
distribution, sub-elements orbiting the center of a large scale structure
at radius $R$ are equivalent to sub-elements orbiting a point mass
with mass $\frac{R}{R_{s}}M_{s}$.

The core volume in a galaxy, containing the concentrated mass of the
supermassive black hole (SMBH), has radius $R_{c}$ related to the
holographic radius of galactic sub-elements that can orbit the center
just outside the core without being disrupted and drawn into the central
black hole. The resulting SMBH mass estimate is $M_{SMBH}(z)=\sqrt{M_{sc}(z)M_{g}(z)}$,
where $M_{g}(z)$ is the total galactic mass and $M_{sc}(z)$ is the
mass of a star cluster mass at redshift $z$ that can occupy a circular
orbit around the SMBH at any radius larger than the holographic radius
of the star cluster, with its holographic screen outside of the SMBH
so it will not be disrupted and drawn into the black hole.

Supermassive black holes can only increase in mass, so the approximate
lower bound on SMBH mass represents an early configuration where matter
within a core radius equal to the holographic radius of the lowest
mass star cluster sub-elements of galaxies is concentrated in the
SMBH. In this configuration, only the smallest (and most numerous)
star cluster sub-elements of galaxies can orbit the galactic center
just outside the core without being disrupted and drawn into the SMBH.
All other star cluster sub-elements must inhabit circular orbits at
distances from the galactic center larger than their holographic radius
to avoid disruption.

The mid-range SMBH mass estimate corresponds to an intermediate case
where matter within a core radius equal to the holographic radius
of the median mass star cluster sub-elements of galaxies is concentrated
in the SMBH. In that situation, star clusters with mass below the
median star cluster mass can orbit the galactic center just outside
the core without being disrupted and drawn into the SMBH. Star clusters
with mass greater than the median star cluster mass must occupy circular
orbits at distances from the galactic center larger than their holographic
radius to avoid disruption.

The approximate upper bound on SMBH mass occurs at a late stage when
matter within a core radius equal to the holographic radius of the
highest mass star cluster sub-elements of galaxies is concentrated
in the SMBH. Then, the full range of star cluster sub-elements of
galaxies can inhabit circular orbits just outside the galactic core
without being disrupted and drawn into the SMBH.

Marleau, Clancy and Bianconi (MCB) \cite{key-20} \emph{et al} summarized
studies of about 6,000 galaxies of different types in a linear equation
relating SMBH mass to total stellar mass of the host galaxy. Total
matter density is 30.8\% of critical density and dark matter is 26\%
of critical density, so this analysis estimates total stellar mass
of galaxies as 15.6\% of total galactic mass. In Figure 1, $\times$
symbols show SMBH mass estimates from the MCB relation based on total
stellar mass of the host galaxy. Square symbols show mid-range SMBH
mass estimates based on median star cluster mass at the appropriate
redshift $z$. Diamond and triangle symbols indicate, respectively,
approximate upper and lower bound SMBH mass estimates based on approximate
upper and lower bound star cluster masses. Overlapping points for
galaxy mass $10^{10}M_{\odot}$ are estimates for galaxies with redshift
$z$ = 0 and $z$ = 0.05. The apparent disagreement for low mass galaxies
is illusory. For example, the MCB relation estimates SMBH mass of
$1.9\times10^{2}M_{\odot}$ for galaxies with total stellar mass $10^{6}M_{\odot}$,
while the actual data (\cite{key-21}, Figure 6) show most SMBH masses
in the range above $10^{3}M_{\odot}$ for galaxies with total stellar
mass $10^{6}M_{\odot}$. SMBH mass estimates in Figure 1 can be compared
with the regression line shown in Figure 9 of Ref. 20 and Figure 6
(right panel) of Ref. 21. Estimates for total stellar mass of $10^{10}M_{\odot}$,
$5\times10^{10}M_{\odot},$ and $10^{11}M_{\odot}$ are results at
$z$ = 0, 0.15, and 0.2 for comparison respectively with blue, green,
and red points at the left, center, and right of the cloud of data
points in Figure 9 of Ref. 20. SMBH estimates at $z$ = 0 for total
stellar mass $10^{6}M_{\odot}$ through $10^{9}M_{\odot}$ should
be compared to data in Figure 6 (right panel) of Ref. 21 that are
generally above the dashed regression line in the figure. For $z$
= 0 to $z$ = 0.25, approximate galactic masses are in the range $10^{6}M_{\odot}$
to $10^{12}M_{\odot}$, and Marleau \emph{et al }data cover this entire
range.

The SMBH mass estimate is also consistent with the estimated mass
of Sagittarius A{*}, the SMBH at the center of our galaxy. The estimated
total dynamic mass \cite{key-22}\cite{key-23} of our Milky Way is
$8\times10^{11}M_{\odot}=1.59\times10^{45}$ g. The corresponding
minimum SMBH mass estimate is $4.8\times10^{39}$g, consistent with
the $9\times10^{39}$g mass estimated for Sagittarius A{*} from astrophysical
measurements \cite{key-24}.

An SMBH can only increase in mass and, within a galaxy, it takes longer
to accumulate the mass in a large SMBH than in a small SMBH. So, this
analysis is consistent with data presented by Merrifield, Forbes and
Terlevich (MFT). The MFT data \cite{key-25} suggest that, for a given
galactic mass, high mass SMBHs are in galaxies ``where the last major
merger occurred long ago'' while low mass SMBHs are in galaxies formed
in more recent mergers. Bluck \emph{et al }\cite{key-26} studied
galaxies with $z<$ 0.2 with stellar mass from $10^{8}M_{\odot}$
to $10^{12}M_{\odot}$. They suggest galaxies with low SMBH mass are
``predominantly star forming'' and galaxies with high SMBH mass
are ``predominantly passive,'' with lower star formation rates than
similar galaxies with low SMBH mass. They find the ``cross-over mass,
where 50\% of galaxies are passive,'' at SMBH mass $\sim10^{7.5}M_{\odot}$.
In this analysis, large SMBH mass (and correspondingly low star formation
rate) should generally occur later in the life of galaxies, as indicated
by Bluck \emph{et al} and MFT. 

Finally, about 40 quasars with $z>6$, containing black holes with
mass $\sim10^{9}M_{\odot},$ have been found so far \cite{key-27}.
Above z =6, a hierarchical self-similar description of large scale
structure is inappropriate, because n(z), the number of levels per
structure, would be less than two. At $z\approx6$, large scale structures
within the Jeans' mass $1.13\times10^{17}M_{\odot}=2.24\times10^{50}$g
would consist of matter in equilibrium with the CMB in the form of
stars with mass between about 300$M_{\odot}$ and the minimum stellar
mass $\sim3.5M_{\odot}$, and the above analysis indicates those structures
should contain SMBHs in the $10^{9}M_{\odot}$range.
\end{doublespace}

\begin{doublespace}

\section{Conclusion}
\end{doublespace}

\begin{doublespace}
None of the results above depend on any arbitrary parameters. In particular,
upper and lower bounds on supermassive black hole mass in relation
to total stellar mass of the host galaxy, consistent with obervations
across four orders of magnitude of black hole mass and five orders
of magnitude of galactic stellar mass, are based only on fundamental
constants and measured cosmological parameters, The fact that no arbitrary
parameters are involved indicates the above analysis provides a coherent
and consistent description of large scale structure in our universe.

Finally, the above analysis applies to a closed universe that is so
large it is nearly flat. Adler and Overduin \cite{key-28} did a careful
analysis of this situation and found that ``observation cannot distinguish
- even in principle - between a perfectly flat Universe and one that
is sufficiently close to flat.'' So, an analysis, based on assuming
a closed inflationary universe containing a finite amount of information,
that accounts for the general features of large scale structture might
serve as an indication that our universe is closed.

\begin{figure}

\caption{\protect\includegraphics{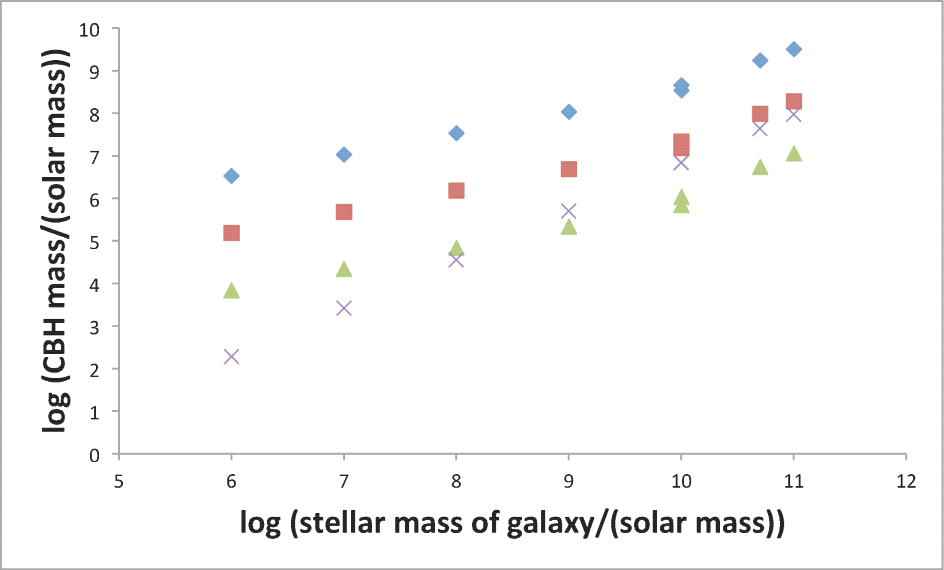}}

\end{figure}

\end{doublespace}

\end{document}